%% file: main.tex
\documentclass[conference]{IEEEtran}
\IEEEoverridecommandlockouts
\usepackage{cite}
\usepackage{amsmath,amssymb,amsfonts}
\usepackage{algorithmic}
\usepackage{graphicx}
\usepackage{textcomp}
\usepackage{xcolor}
\usepackage{multirow}
\usepackage{hyperref}
\usepackage{booktabs}
\usepackage{rotating}
\usepackage{tabularx, makecell}
\usepackage{array}
\usepackage{pifont}

\newcolumntype{L}{>{\centering\arraybackslash}m{3cm}}
\def\BibTeX{{\rm B\kern-.05em{\sc i\kern-.025em b}\kern-.08em
    T\kern-.1667em\lower.7ex\hbox{E}\kern-.125emX}}
\begin{document}

\title{Using Hypotheses as a Debugging Aid}

\author{\IEEEauthorblockN{Abdulaziz Alaboudi}
\IEEEauthorblockA{\textit{George Mason University} \\
Fairfax, Virginia, USA \\
aalaboud@gmu.edu}
\and
\IEEEauthorblockN{Thomas D. LaToza}
\IEEEauthorblockA{\textit{George Mason University} \\
Fairfax, Virginia, USA \\
tlatoza@gmu.edu}
}

\maketitle
\begin{abstract}
As developers debug, developers formulate hypotheses about the cause of the defect and gather evidence to test these hypotheses. To better understand the role of hypotheses in debugging, we conducted two studies. In a preliminary study, we found that, even with the benefit of modern internet resources, incorrect hypotheses can cause developers to investigate irrelevant information and block progress. We then conducted a controlled experiment where 20 developers debugged and recorded their hypotheses. We found that developers have few hypotheses, two per defect. Having a correct hypothesis early strongly predicted later success. We also studied the impact of two debugging aids: fault locations and potential hypotheses. Offering fault locations did not help developers formulate more correct hypotheses or debug more successfully. In contrast, offering potential hypotheses made developers six times more likely to succeed. These results demonstrate the potential of future debugging tools that enable finding and sharing relevant hypotheses.
\end{abstract}

\begin{IEEEkeywords}
Debugging, hypotheses, fault localization
\end{IEEEkeywords}

\input{sections/1.introduction.tex}

\input{sections/2.background.tex}

\input{sections/3.preliminary.tex}
\input{sections/4.Methods.tex}
\input{sections/5.Results.tex}

\input{sections/6.limitation.tex}
\input{sections/7.Discussion.tex}

\section*{Acknowledgments}
We would like to thank the participants for their time. This research was funded in part by NSF grant CCF-1703734 and CCF-1845508. The first author is supported in part by a King Saudi University Graduate Fellowship.

\bibliographystyle{IEEEtran}
\bibliography{main}
\end{document}

%% file: sections/1.introduction.tex
\section{Introduction}

Debugging has long been a focus of software engineering research, encompassing studies of the debugging process as well as the creation of numerous techniques to more effectively support it \cite{Myers2012ArtOfTesting, Zeller2005, LaToza2010ReachabilityQuestions, Ko2006aDebuggingByAskingQuestions, Lawrance2013IFT,MarkWeiser1984ProgramSlicing, Ko2008DebuggingReinvented}. Key to the process of debugging are hypotheses. A debugging hypothesis is a verifiable speculation about the possible cause of the incorrect behavior \cite{Zeller2005, Ko2007InformationNeeds, Perscheid2017}. 
Developers build mental models of the program by asking questions about the incorrect behavior of the program,  hypothesizing possible causes, and collecting information to test them \cite{Ko2007InformationNeeds, Roehm2012,Sillito2008}. 
For example, a developer who sees a search feature fail might ask, "Why did the search not return the correct answer?". She might then hypothesize that it was caused by an incorrect comparison in its implementation of string matching. From this hypothesis, she might gather evidence to test it, searching for locations related to string matching and using the debugger to gather information about the run-time state to determine if each step in the string matching algorithm is correct\cite{Perscheid2017, Layman2013}. 

Unfortunately, developers often formulate incorrect hypotheses, resulting in wasted time gathering evidence and looking at irrelevant code that ultimately does not lead the developer closer to the true defect \cite{Ko2006}. When they fail to find a correct hypothesis, developers often look for help, mitigating the need to generate their own hypotheses. Seeking help from an experienced coworker is one way to find the correct hypothesis. Unfortunately, developers may not always find their coworkers available \cite{Ko2007InformationNeeds}. One might also expect that, given the wealth of developer information available on the internet, finding hypotheses might be easy.
To explore this, we conducted a small preliminary study in which we observed three professional developers working in three open-source projects. We found that developers often got stuck because they lacked correct hypotheses or had an insufficiently precise hypothesis. Lacking a correct hypothesis, developers formulated search queries beginning from incorrect output (e.g., error messages) or an insufficiently specific hypothesis (e.g., example of API usage based on the hypothesis that the API is being used incorrectly). This resulted in irrelevant information that did not lead to a fix, wasting further time.

Despite their centrality to debugging \cite{Perscheid2017}, many important questions remain unanswered about the role of hypotheses in debugging. 
It is unclear how hard it is to formulate correct hypotheses and how closely hypotheses are tied to developers' debugging performance. And, in situations where developers lack hypotheses, questions remain about how debugging aids might assist developers in finding hypotheses, such as suggesting potential fault locations to investigate or directly offering developers potential hypotheses. 
   

To fill this gap, we investigated three research questions:

\begin{itemize}
    \item[\textbf{RQ1}] How hard is it to formulate correct hypotheses? Does formulating correct hypotheses predict debugging success?

    \item[\textbf{RQ2}] Does  offering developers fault  locations help developers to form correct hypotheses and debug more successfully?
    
    \item[\textbf{RQ3}] Does offering developers potential hypotheses help developers debug more successfully? 
\end{itemize}
    
We conducted a lab study in which 20 developers worked to debug defects in three small programs taken from Stack Overflow. We chose to focus on API-related defects, as studies suggest these can be challenging to debug \cite{CokerQualitative2019}. 
To observe the process of how developers formulate hypotheses during debugging, we organized the debugging tasks into three \textit{stages} and asked developers to write down their hypotheses at each stage. At each stage, developers were given access to more information, including the bug report and user interface (stage 1), source code and related documentation (stage 2), and the ability to run and edit the code (stage 3).  
This enabled us to observe hypotheses before developers were able to test and discard them. 
Additionally, some participants were given access to either potential fault locations or hypotheses.

We found that participants struggled to formulate correct hypotheses. Per defect, the median number of hypotheses participants formulated across all stages was two. In the first stage, only 11\% of hypotheses were correct. While this increased in stage two and three, overall only 39\% of reported hypotheses were correct. However, for developers that did discover the correct hypothesis by stage two, their chance of ultimately succeeding in debugging increased by a factor of five. Offering developers fault locations did not significantly help developers to formulate a correct hypothesis nor enable the developer to debug any more successfully. In contrast,  developers who received potential hypotheses were six times more successful in fixing defects. 

These results suggest the potential value of new types of debugging aids that offer potentially relevant hypotheses to developers. We discuss the implications of our findings for future tools that enable developers to share, find, and use debugging hypotheses.

%% file: sections/2.background.tex
\section{Background}
Early studies of code comprehension and debugging investigated the role of code understanding and developers' mental models. These studies found that developers formulate hypotheses during programming tasks \cite{Brooks1983, Letovsky1987,VonMayrhauser1996, Jeffries1982a, Littman1987}. Brooks  introduced a theory in which developers form hypotheses about program behaviors while comprehending code\cite{Brooks1983}. Letovsky proposed that hypotheses (``conjectures'') were answers to questions developers ask while comprehending code\cite{Letovsky1987}. 

Several prior studies have observed developers formulate and test hypotheses. Studies have found that developers ask questions about incorrect output and form hypotheses about its cause \cite{Ko2007InformationNeeds,Ko2006aDebuggingByAskingQuestions}. Developers then test these hypotheses by examining code locations and inspecting program state \cite{Perscheid2017}. Robin investigated how novices and experts differ in their debugging process and found that experts were more likely to formulate correct hypotheses and successfully map them to the related code \cite{Jeffries1982a}. Other work has tried to avoid modeling debugging hypotheses explicitly by instead modeling developers' actions, which may be easier to observe. According to information foraging theory, developers navigate between methods ("patches") based on method identifiers which offer scent that hint at their proximity to the fault location ("prey")~\cite{Lawrance2013IFT}.

Professional developers largely depend on the ``intuitive method'' of forming and testing hypotheses while debugging \cite{BohmeFSE-DebuggingHypotheses, Perscheid2017}. Unfortunately, developers may formulate incorrect hypotheses, leading them to inspect irrelevant code \cite{Ko2006}. As developers waste time testing incorrect hypotheses, many tools have been proposed which envision entirely bypassing the need for a developer to formulate a hypothesis. Debugging is often framed as a problem of fault localization, where a developer begins with a fault and must identify the statement responsible \cite{Vessey1985,Myers2012ArtOfTesting}. Many tools and techniques have been built to help developers in searching for fault locations. For example, program slicing tools  often display a ranked list of potentially faulty statements, shrinking the search space of potentially faulty statements developers must presumably consider \cite{MarkWeiser1984ProgramSlicing, DeMillo1996, Zhang2006, XiangyuZhang2003}. However, automatic fault localization tools are typically evaluated in their performance of reducing the search space rather than in their ability to improve developers overall debugging performance. Implicit in this work is the assumption that reducing the set of faulty statements a developer must consider will necessarily improve debugging performance. One study tested this assumption directly through a user study and found that fault localization techniques do not always help developers debug more effectively \cite{Parnin2011AreAutomatedDebugging}. 

A common practice when developers confront faults for which they have no hypothesis is to consult external resources. In co-located development teams, developers may consult an experienced developer. However, studies have found that developers may not always find an experienced coworker available \cite{Ko2007InformationNeeds, LaToza2006MentalModel}.  Developers may also search the internet for potential hypotheses, posting questions and examining responses to prior questions in community resources such as Stack Overflow \cite{Yang2013, Vasilescu2014}.
Success depends on several factors, including asking sufficiently precise questions and providing enough information about the problem\cite{Treude2011,Mamykina2011,Asaduzzaman2013}. 

While prior work has examined how developers form and test hypotheses, little work has systematically examined how a range of factors impact the generation and use of hypotheses in the debugging process. To our knowledge, we are the first study to ask developers to repeatedly report their debugging hypotheses and examine the relationship between situational factors, the generation of hypotheses, and debugging success. We offer the first evidence about how offering fault locations or a set of potential hypotheses might help developers in formulating and testing hypotheses.

%% file: sections/3.preliminary.tex
\section{Preliminary Study: Searching for help}
To better understand the challenges developers may face in formulating hypotheses, given the opportunity to use modern Internet resources, we conducted a small preliminary study. To observe developers, we chose to use live-streamed programming videos, where developers live-stream their programming activities for other developers to watch. \cite{AlaboudiVLHCC2019}. In live-streamed programming, developers live-stream their work on real-world development tasks fixing issues and implementing features for open source projects. During the live-streamed session, developers keep their viewers engaged by frequently thinking aloud, articulating intentions, goals and hypotheses, which make these videos a valuable data source for observational studies \cite{Alaboudi2019CHASE}. After finishing streaming their work, developers archive the videos on online platforms such as YouTube.  We selected three videos of developers (D1, D2, and D3) working on different open source projects hosted in GitHub for 177, 85, and 104 minutes, respectively. The selection process followed a systematic approach. We first searched for videos posted in developer communities such as the r/WatchPeopleCode subReddit. We then checked if the video showed developers working on an open source project by searching for the project in GitHub. Finally, we skimmed the video to determine if it included the developer searching the Web while debugging. Our replication package contains the data for this study \cite{replication}. 

\begin{figure}
    \includegraphics[ width=\linewidth, keepaspectratio, clip]{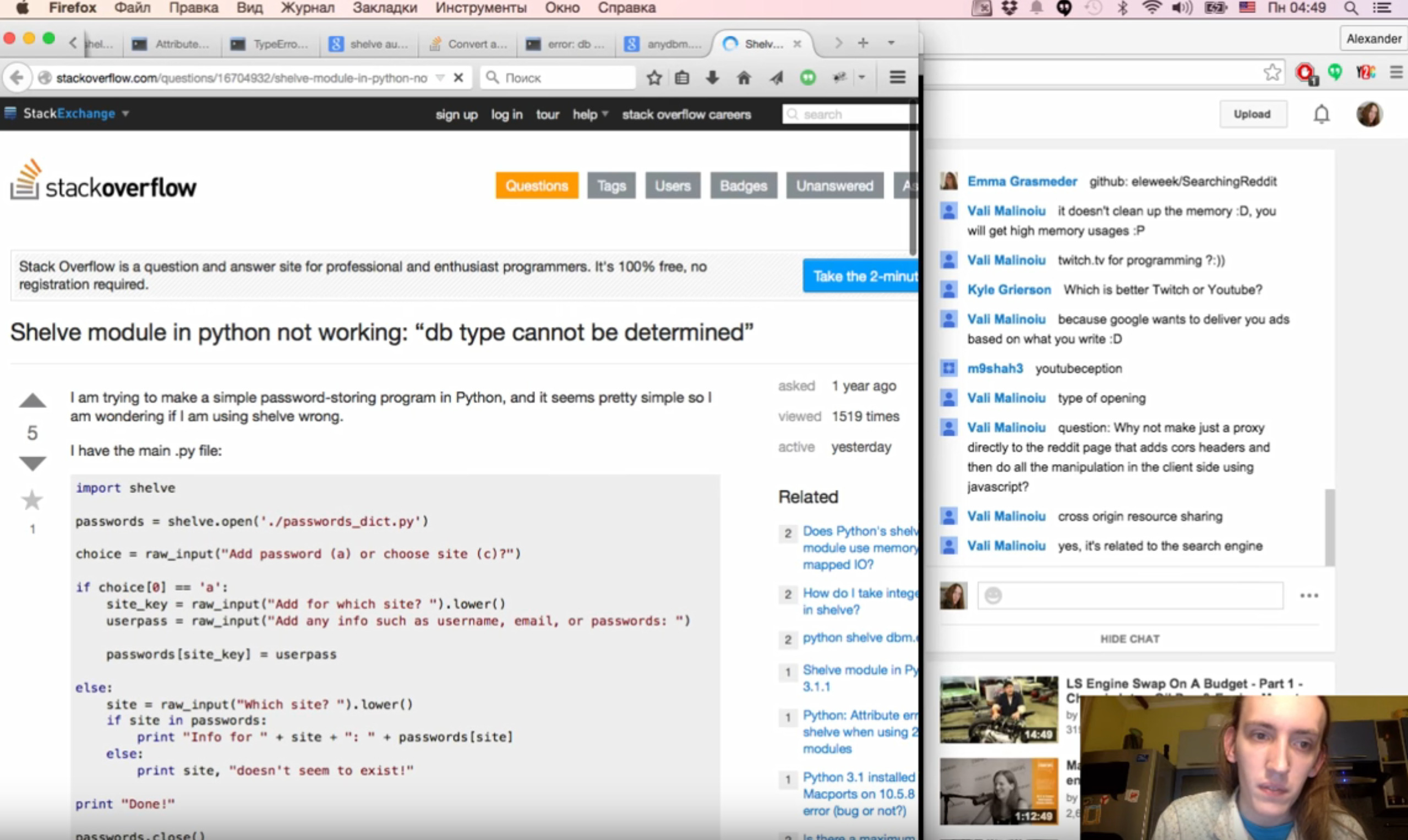}
    \caption{ \label{fig:SOF}Faced with an error message, D1 read a Stack Overflow post. }
  \end{figure}

D1 was building a search engine for Reddit in Python. During his three hour programming session, he encountered a defect related to his use of the shelve API. He hypothesized that he ``maybe needs to specify the file type`` when calling the API. Using this hypothesis, he searched the Internet and started reading the official documentation for a few minutes, stopping reading as there was no API to specify the file type. Unable to hypothesize, he switched to Google and pasted the error message in the search field. He started browsing a related Stack Overflow post (Figure \ref{fig:SOF}) that suggested potential solutions. He decided to try a solution, stating that ``this looks like the problem I am having now''. He copied the suggested fix and pasted it into his code, but the error still occurred. Blocked from further progress in his debugging task, he decided to end the session. In his next video, he discovered the bug was due to an incorrect parameter to one of the APIs calls. He did not consider this possibility in formulating his search query. 

In his 85 min programming session, D2 worked to build a face detection program using C++ and openCV APIs. While trying to compile the program, he received an error message of ``undefined reference to cv''. His first hypothesis was that the defect was due to incorrect usage of the standard I/O APIs. However, after changing it to a simple print statement, the code still did not compile. He then copied part of the error message and searched the Internet. He spent around 20 minutes testing multiple hypotheses suggested on Stack Overflow, such as checking for typos and changing compilers. However, none were correct. Two of the developers watching his live-stream offered the hypothesis that the defect was due to an issue in the way the OpenCV library was linked at compile time. After searching with just three words, ``link to openCv'', he was able to find relevant information and successfully fix the defect.

D3 worked to implement a confirmation dialog in JavaScript using the Electron APIs. He was unable to make the dialog's cancel and confirm options work correctly. He searched the Internet for documentation and code examples on how to use specific APIs. After looking at an example, he hypothesized that the bug is related to the async APIs. "Oh, it has async APIs. Sorry, I am just a noob.''. However, changing the implementation to use async APIs did not fix the defect. He kept changing the return value of a callback function and inserting log statements to answer many ``what if'' questions.  After debugging for half an hour, he reported that he could not fix the defect, stopping work and seeking help later from other developers. We investigated the pull request he was working on and found he was able to fix the defect the following day. From the commit history, it appears that the defect was related to a missing API call unrelated to the documentations or APIs he browsed in the video.

Together, these observations suggest the centrality of hypotheses to debugging. Even with the benefit of modern Internet resources, developers often struggled to know where to start. Lacking a correct hypothesis, developers formulated queries hoping to discover something relevant and looked at information that often led nowhere. These findings suggest the need for tools that help developers better search for and acquire hypotheses while debugging. State of the art debugging tools such as WhyLine attempt to entirely avoid hypotheses by, for example, enabling developers to systematically trace incorrect output backwards across control and data flow. However, these techniques may be ill-suited for situations involving interactions with complex third-party APIs, such as in our preliminary study, where developers cannot or do not wish to follow data flow through framework internals. This suggests the need for new forms of debugging aids that better support the process of hypothesis formulation and testing.   




%% file: sections/4.Methods.tex
\section{Controlled study: Investigating Hypotheses in debugging}

We conducted a within-subjects study in which 20 participants completed a series of three debugging tasks to fix API misuse defects. In this section, we describe our study and the choices we made in its design. 

\subsection{Study Design}
In order to observe debugging hypotheses, we chose to divide our study into three separate stages. This enable us to understand the time and resources required to form correct hypotheses as well as avoid the possibility that developers might test and reject hypotheses before reporting them.

In each stage, participants were instructed to report their hypotheses and the nessary steps to test them. In the first stage, participants were able to execute the program and observe its behavior through GUI or console interactions, without access to the source code or documentation.  In the second stage, participants were given access to related documentation and a snapshot of the source code, but without the ability to edit or execute the program. Participants were asked in this stage to submit new hypotheses and edit hypotheses submitted in stage one. In the third stage, participants had the ability to edit and execute the program. They were asked to continue forming new hypotheses or edit existing ones, if needed, and use any hypotheses they had to fix the defect. To investigate the effects of offering debugging aids, participants were assigned to one of three conditions for each task: a control condition (no aid), a fault locations condition, and an potential hypothesis condition. Debugging aids were offered to participants when they first were given the source code in stage two. Participants did not have access to the Internet throughout the study. Table \ref{fig:debuggingStages} lists the goals and provided resources for each stage. 

\begin{table}
\caption{ goals and resources for each stage. The gray marks in the bottom row indicate that a debugging aid was offered to participants in two randomly assigned  tasks.}
\begin{tabular}{llccc}
    \toprule
                     & & \multicolumn{1}{l}{\begin{tabular}[c]{@{}l@{}}Stage 1\\ 7 min\end{tabular}} & \multicolumn{1}{l}{\begin{tabular}[c]{@{}l@{}}Stage 2\\ 10 min\end{tabular}} & \multicolumn{1}{l}{\begin{tabular}[c]{@{}l@{}}Stage 3\\ 20 min\end{tabular}}   \\ 
    
                        \cline{2-5}
                         \addlinespace 

                            \multirow{2}{*}{\begin{turn}{90}Goals\end{turn}}     
                            & Formulating hypotheses& \ding{51}& \ding{51}& \ding{51}\\
                            &Fixing the fault & & & \ding{51}\\
                            \addlinespace 
                            \multirow{6}{*}{\begin{turn}{90}Resources\end{turn}} & Fault report & \ding{51} & \ding{51}& \ding{51}\\
                           & Execution environment & \ding{51}& &\ding{51}\\
                           & Picture of the correct output & \ding{51} & \ding{51}& \ding{51}\\
                           & Documentation & & \ding{51} & \ding{51}  \\
                           & Source code &  & \ding{51} & \ding{51}  \\
                           & Debugging aids &  & {\color{gray}\ding{51}} & {\color{gray}\ding{51}} \\
                           \bottomrule
\end{tabular}
\label{fig:debuggingStages}
\end{table}

\subsection{Tasks}
To encompass a variety of modern APIs, we selected three programs taken from Stack Overflow questions that are related to different Web technologies.  Each program contained a defect that prevented it from behaving as intended. We offered participants the reported Stack Overflow question as a prompt and the code snippet as the provided program. To ensure that participants understood the intended behavior,  participants received a picture or animated gif of the expected output.

In the first task, participants worked with a program that used the \textit{DOM} (Document Object Model) API to change the visibility of an HTML element using the ``visibility'' DOM property. The program consists of 37 lines of HTML, CSS, and JavaScript code. The defect concerned an incorrect use of the common JavaScript API ``getElementsByClassName''. This API returns an array of DOM elements. The implementation contained a defect where it instead treated the return value as a single DOM element, which can be fixed by iterating over the returned collection and setting the visibility for each element.

In the second task, participants worked with a program built using \textit{React}, one of the most commonly used UI frameworks\footnote{https://insights.stackoverflow.com/survey/2019/}, containing 36 lines of HTML and JavaScript.   The intended behavior of this program is to switch images on a click. This can be achieved by successfully registering a callback for the onClick event. However, the implementation of the registration contains a defect, resulting in the onClick event handler generating an undefined error.

As animation is common in Web applications, we selected for the third task an animation program containing 45 lines of HTML, CSS, and Javascript. The intended behavior is to move a figure from one position to another using \textit{jQuery} animation APIs, which it fails to achieve. To fix this defect, the correct position value has to be passed to the jQuery API.

\begin{figure}
    \includegraphics[width=1\columnwidth,keepaspectratio, clip]{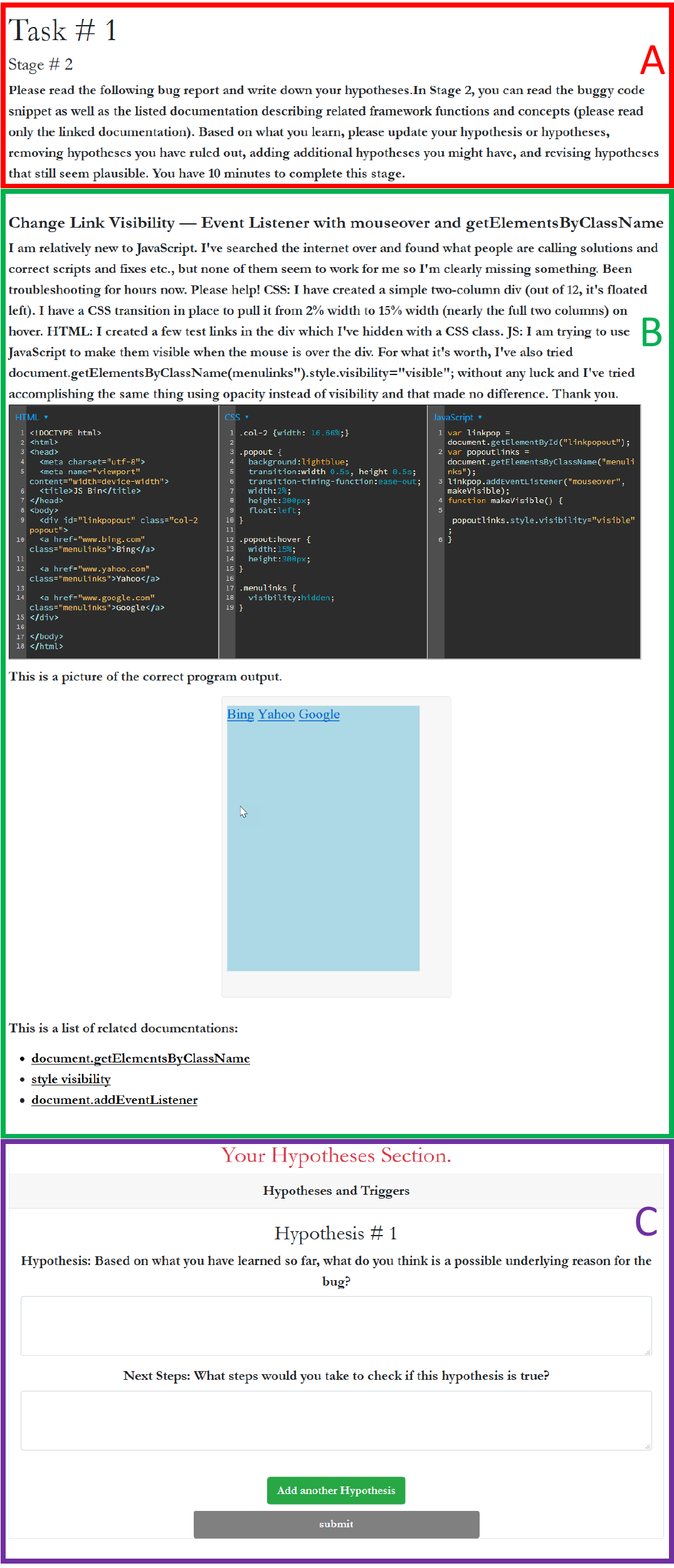}
    \centering
    \caption{\label{fig:studyExample} In the study environment, participants completed each debugging stage in a separate page. (A) The top of each page describes the goal and time limit for the current stage. (B) The middle section offers a debugging environment,  with the applicable resources for the stage. (C) The bottom section  enables participants to write down their hypotheses as well as offering a debugging aid, if applicable.}
\end{figure}

\begin{figure*}
    \includegraphics[width=1\textwidth]{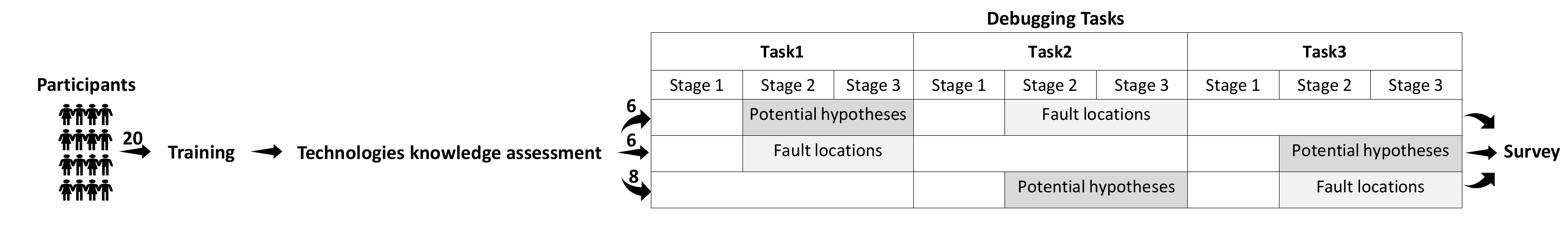}
    \centering
    \caption{\label{fig:studyDesign} During the study, participants first completed a training task and a technology knowledge assessment. Participants were then randomly divided into three groups. A Latin square design was used to assign debugging aids to debug tasks, controlling for potential order and task effects.}
\end{figure*}

\subsection{Debugging Aids}
We intended for the debugging aids to be representative of aids that might be generated by current and future tools, as we were interested in studying the impact of these debugging aids without any assumptions of how such aids were generated. In addition to studying the impact of potential hypotheses as a debugging aid, we studied the impact of offering fault locations, as commonly used in research prototypes offering automated fault localization. Participants received either two potential hypotheses or two fault locations. To control for the performance of tools generating these aids, we simulated the output that one of these tools would generate. In both tasks, participants received one correct and one incorrect potential hypotheses or one correct and one incorrect fault locations with the incorrect one always listed first, simulating the behavior of a tool with high, but not perfect, precision.

Each potential hypothesis consisted of two parts: the hypothesis itself and steps describing how to test the hypothesis. We extracted the correct hypotheses and the steps on how to test them from the corresponding correct answer in the Stack Overflow question. Sine the potential hypotheses are representative of a debugging aid offered by a tool, the hypotheses are generalized to be applicable for different program contexts. We summarized the answer and then removed any references to implementation details in the hypothesis. For the incorrect hypotheses, we searched Stack Overflow for a question that described similar symptoms but with a different underlying cause. To generate the incorrect and correct fault locations, we used the potential hypotheses we extracted for each task and mapped these hypotheses to locations in the source code. These locations were then presented to participants. Table \ref{tb1:hypothesesTable} shows the offered potential hypotheses.

\subsection{Participants}
We recruited 25 participants (15 male and 10 female) from a graduate course in software engineering at our institution.
Five of these participants were used to pilot the study materials and are not described further. As typical in industry,  participants reflected a wide range of experience levels, including many who worked full time as a software developer.

To control for experience level, we measured the programming expertise of each participant.
To measure programming experience, we asked participants to report their number of years of programming experience in industry. Measuring the level of technology knowledge  is more challenging, as relying on participants' self-assessments may introduce bias in the results \cite{Baltes2018}. Thus, we instead chose to develop a technology knowledge assessment that can be completed quickly. We developed technology assessments with nine multiple-choice questions. These questions assessed knowledge of fundamental to advanced concepts in each of the technologies used in the study's tasks. Each question had five possible answers, including a "I do not know" choice.  To avoid any potential learning effects, we did not ask participants about any concepts or APIs that directly related to the defects in the study tasks. Participants received one point for each correct answer.

Participants ranged from 0-20 years of industrial experience (median = 2) and, on the 9 point scale of the technical knowledge assessment, scored from 1-8  (median = 4).

\subsection{Procedure}
To conduct the study, we constructed a purpose-built study environment website that administered the technology assessment questions, the debugging tasks, and a short post-task survey. The study environment contained all information about the debugging tasks and enabled participants to view, edit, and run the code, depending on the debugging stage. The study environment collected the technical assessment answers, fault patches, stages time, and survey responses. During each debugging stage, participants were prompted to enter one or more hypotheses. Figure \ref{fig:studyExample} depicts an example of a stage two page in the environment. The study environment, study materials, and other data are all publicly available \cite{replication}.

We conducted our study in three small-group sessions, each lasting approximately 2.6 hours.  Participants in each session were gathered in a room and asked to access the study environment from their computers. 
To familiarize participants with the study environment, participants first completed a training task that consisted of a debugging task containing each of the three stages. One of the authors was available to answer questions. Participants next completed the nine questions in the technical knowledge assessment and then began each stage for each of the three debugging tasks.
Participants were given a time limit of seven minutes in stage one, ten minutes in stage two, and 20 minutes in stage three. Participants who ran out of time in a stage were automatically advanced to the next stage. Finally, participants completed a post-task survey asking them to report the number of years they have been programming professionally and describe whether they found each of the debugging aids to be helpful. Participants had up to 3 hours to complete the study. Figure \ref{fig:studyDesign} illustrates the study design.

\subsection{Data Analysis}
 
To score each hypothesis, we labeled a set of keywords that reflecting the underlying cause in the correct debugging hypotheses aid. Participants' hypotheses were scored as correct if they described a concrete and verifiable reason for the cause of the defect using  the same or similar keywords as the correct debugging hypotheses aid. For example, for a defect that was caused by passing a string instead of a number to a third party library, a hypothesis such as ``the defect is caused by a third party library'' was considered too vague to correctly describe the cause of the defect while ``the defect was caused by an incorrect parameter value passed to a third party library'' was considered concrete and correct. Two authors independently coded participants' hypotheses as correct or incorrect, resulting in a Cohen's Kappa value of 0.76, reflecting substantial agreement \cite{Richard1977}. The disagreements were resolved through a discussion between the authors. 

To answer each research question, we built a series of regression models. To control for the effects of expertise \cite{Vessey1985} and task \cite{Perscheid2017} in our models, we included years of programming experience, technical knowledge, and task as control variables in each model. Adding a control variable for task controlled for variance between tasks in difficulty, where task three had a 90\% success rate and tasks one and two had 55\%.

%% file: sections/5.Results.tex

\section{Results}
 
\subsection{How hard is it to formulate correct hypotheses? Does formulating correct hypotheses predict debugging success?}

We first investigated the difficulty of formulating  correct hypotheses without any debugging aids. We analyzed data from the 20 debugging sessions in the control condition. Participants formulated 27 hypotheses (Min 1, Max 3, median 1) in stage one. Of these 27 hypotheses, only three were correct (11\%). In the second stage, participants added eight new hypotheses and edited six (total 35, Min 1, Max 3, median 1), which increased the number of correct hypotheses to nine (25\%). In the last stage, participants added three new hypotheses and edited four (Total 38, Min 1, Max 4, Median 2), which increased the number of correct hypotheses to 15 (39\%). Although our study was designed to encourage participants to formulate as many hypotheses as they can, the median number of reported hypotheses for each participant was two. This suggests that not only the process of formulating a correct hypothesis is challenging, but also the process of formulating any reasonable and not necessarily correct hypothesis.  Table \ref{tb1:hypothesesTable} lists examples of participants' hypotheses in stage two. The complete list of hypotheses is available online \cite{replication}

We next investigated the relationship of hypotheses to debugging success, irrespective of how the hypotheses were obtained. We analyzed the data from the 60 debugging sessions from participants in the three conditions. We chose a dependent variable of correct or incorrect hypothesis and built a logistic  regression  model. The model shows that participants who reported a correct hypothesis at stage two  were 5 times more likely to fix the defect (p = 0.01). 

However, having the correct hypothesis usually but not always leads to success in debugging. Overall, participants failed in 20 out of 60 cases to fix the defect. Six of these 20 cases involved participants with a correct hypothesis at stage two who did not successfully fix the defect. To investigate these cases, we analyzed participants' survey responses and code edits. Their responses suggest that, while they understood the underlying cause of the fault, they did not know how to correctly implement a fix. One participant, for example, realized that the fault in the first task was due to a DOM API returned a collection of DOM elements instead of a single DOM element, but tried to fix the fault by changing another irrelevant part of the code:

``I tried changing the name of the item to which the event handler was added, but it still [did] not make the div appear. The links are not being set to visible because they are in an array.''

\begin{table*}[ht]
    \centering
        \caption{The potential  hypotheses given to participants and examples of participants' correct  and incorrect hypotheses at stage two.}
        
    \begin{tabular}{m{3mm} p{40mm} p{40mm} p{40mm} p{40mm}}
         \toprule
        Task&Correct potential hypotheses & Incorrect potential hypotheses & Examples of participants’ correct hypotheses& Examples of participants’ incorrect hypotheses \\
         \midrule
        \multirow{3}{*}{1} &
        \textbf{Hypothesis}: You are using the wrong DOM API 
        or not using getElementsByClassName correctly. \newline
        \textbf{How To Test}: Check if you are using the correct DOM API. Also remember: getElementById returns an HTML element that has the same id, get ElementsByClassName returns a HTML Collection that you have to iterate over.
        &
         \textbf{Hypothesis}: You are not using the correct CSS property. \newline
        \textbf{How To Test}: Check if you are using the correct CSS property to change the visibility of an element. Example: visibility: 'visible' or 'hidden', or  display: 'none' or 'block'.
         & 
         \textbf{Hypothesis}: style changed is wrongly applied to a collection variable.
        \newline
        \textbf{How To Test}: Change the style application to a loop over all elements in collection.
        &
           \textbf{Hypothesis}: In the JS code the HTML Div Element's selector (id, class, or xpath) is not properly defined. There is an error in the selector that is causing the error.
        \newline
        \textbf{How To Test}: Check the HTML code and the JS code to see if the JS code is using the correct selection. \\
         \midrule

        \multirow{3}{*}{2} &
        \textbf{Hypothesis}: You are not binding the scope of 'this' for the callback.
             \newline
        \textbf{How To Test}: Check that you are binding the callback for the click event with 'this’. For example: callback.bind(this).

         & 
          \textbf{Hypothesis}: You are not creating the initial state.
             \newline
        \textbf{How To Test}: Check if you have created initial state in the constructor. For example: this.state = \{\}.

         & 
         \textbf{Hypothesis}: The handleClick method has not been bound to the React component instance, and so calling this.handleClick in the event handler set the calls "this" reference to an object other than the component, and so the setState method is not called on the right object.
        \newline
        \textbf{How To Test}: Bind the handleClick method to the instance and try running it again.
        &
           \textbf{Hypothesis}: this.state is not being defined properly. Since it is being done in constructor, it is causing this.setState to be null.
        \newline
        \textbf{How To Test}: check the correct way to define state.\\
         \midrule

                \multirow{3}{*}{3} &
        \textbf{Hypothesis}: You are not setting the new animation position correctly.
 \newline
        \textbf{How To Test}:Check that the new position value is set correctly and in pixels. For example: 15px.
        &
              \textbf{Hypothesis}:You are not assigning a callback to the click event.
 \newline
        \textbf{How To Test}:Check if you have a callback function to the click event. JQuery has this pattern: \newline \$('\#htmlId').click(callback)
         & 
         \textbf{Hypothesis}: The animation CSS under the else statement keeps the object at the same place as the if statement, which is why the object doesn't move
        \newline
        \textbf{How To Test}: Try changing the position of object in CSS in the else statement around to see if the animation starts changing.
        &
           \textbf{Hypothesis}:The clicked boolean property is being changed at an incorrect location that is causing the if/else to malfunction. \newline
        \textbf{How To Test}: remove the line 12 and change line 20 to be changed = !changed; \\
        \bottomrule
        
    \end{tabular}
    \label{tb1:hypothesesTable}
\end{table*}

\subsection {Does  offering  developers fault  locations help developers to form correct hypotheses and debug more successfully?}

To test if offering developers fault locations increases their ability to formulate correct hypotheses at stages two and three, we built a logistic regression model with the correctness of the hypotheses as the dependent variable and the availability of fault locations as the independent variable. We found no evidence that by offering fault locations participants formulate more correct hypotheses, either in stage two (p = 0.4125) or stage three (p = 0.1320). Overall, 45\% of the participants in stage two and 60\% in stage three from the fault locations group submitted the correct hypotheses compared to 45\% and 75\% participants, respectively, from the control group. Offering potential fault locations did not significantly help participants fix the defect (p = 0.6710). 

To explore why providing potential fault locations did not yield observable benefits, we reviewed participants' survey responses about how they made use of the fault locations during the debugging tasks. Some participants stated that they felt that having fault locations helped them to narrow their search space. However, several indicated that they needed additional insight beyond the fault locations to understand the underlying cause of the defect.

\begin{quote}
``I looked at the buggy line help, [but] I was still not sure how to fix the code.  However, at least I could focus just on those lines instead of all of the rest of the application.''
\end{quote}

\noindent Another participant reported that the fault locations might help in situations where she has more knowledge about what might cause the fault.

\begin{quote}
``It seemed helpful only when I was slightly confused. On conditions when I had no clue, it could have been more helpful by providing more feedback other than just [the] line number, like perhaps the correct syntax when it is wrong or incomplete.''
\end{quote}




\subsection {Does offering potential hypotheses help participants debug more successfully?}

We investigated the impact of directly offering potential hypotheses on debugging success. Table \ref{tb4:generalTable} summarizes the results.
We found that participants who received potential debugging hypotheses were six times more likely to fix the fault compared to the other groups (p = 0.0388). Surprisingly, we found that offering potential hypotheses was a stronger predictor for debugging success than participants' years of expertise (table \ref{tb3:debuggingAids}). 

The survey indicates that participants wanted the potential hypotheses aid to be more helpful in locating the source code related to the hypothesis and to include an example.

\begin{quote}
``[Potential hypotheses]  weren't usually very descriptive, and/or didn't show an example. This would have made them more helpful''. 
\end{quote}

\begin{quote}
``[Potential hypotheses] gave me a good suggestion to find the bugs. They did not tell me the specific position, and after I thought for a while, I could find it out.'' 
\end{quote}

\begin{table}[ht]
    \centering
        \caption{Task success and median time to success in minutes for each task by condition.}
    \begin{tabular}{lp{6mm}p{6mm}p{6mm} p{6mm}p{6mm}p{6mm}}
    \toprule
        \multirow{2}{*}{Condition} & \multicolumn{3}{c}{Success} &  \multicolumn{3}{c}{Median Time} \\
         &Task1 & Task2 & Task3 & Task1 & Task2 & Task3\\
         \midrule
       Control & 4/8 & 2/6 & 5/6  & 28 & 13.5 & 14.3\\
      Fault locations & 2/6 & 4/6 & 7/8 & 16.6 & 25.3 & 21.6 \\
      Potential hypotheses & 5/6 & 5/8 & 6/6 & 24 & 18.1 & 12 \\
      \bottomrule
    \end{tabular}
    \label{tb4:generalTable}
\end{table}

\begin{table}
    \centering
\caption{An ordinal logistic regression to evaluate the impact of two debugging aids on debugging success.}
\begin{tabular}{l c c c l}
\toprule
 variables & Odd ratio & SE $\beta$ & Wald & Sig. (p) 
 \\
         \midrule

Fault locations &1.4& 0.77& 0.4& 0.64 \\
Potential hypotheses& 6.24& 0.88& 2& 0.03*\\
Years of experience& 1.14& 0.06& 2& 0.04* \\
Technology knowledge & 2.19 & 0.43 & 1.81 &0.07. \\
\bottomrule
    \end{tabular}
    \label{tb3:debuggingAids}
\end{table}
These results suggest that potential hypotheses  helped to set participants in the right direction to fix the defect and greatly increased their chance of success. However, we did not find any significant differences in debugging time (f = 0.51, p = 0.61). To fix the defect,  participants still needed to understand the hypothesis, find how it related to the defect, and read additional information such as documentation. Participants who succeeded in the control, fault locations, and potential hypotheses conditions spent on average 5\%, 17\%, and 6\% of their time reading documentation, respectively, with no significant differences between conditions (f = 1.7, p = 0.19).



%% file: sections/6.limitation.tex

\section{Limitations and Threats to Validity}

Our investigation of debugging hypotheses has several potential threats to validity. One potential threat to validity is the context in which participants worked during the study. We asked participants to use our study environment throughout the study. To control for potential variation in participants' search strategies in using external resources, we asked participants to not use the Internet and only use the resources offered to them. While this does not represent the typical context of work environments, we closely simulated how web developers work by embedding a well-known editor\footnote{https://jsbin.com} in the study environment and gave participants the best documentation we found related to the debugging tasks. 

An additional potential threat to validity is in our selection of debugging tasks. All of the tasks involved debugging simple Web applications. Offering fault locations might benefit developers more for larger programs. However, each task represented a real debugging situation reported in Stack Overflow after a developer struggled and needed help. 

In our study, we extracted developers' hypotheses by asking them to enumerate as many hypotheses as possible before testing them. Although this may not represent how developers debug using the scientific debugging strategy\cite{Zeller2005}, it forces developers to think through actions before taking them,  closer to the popular practice known as rubber duck debugging \cite{huntpragmatic}.

%% file: sections/7.Discussion.tex
\section{discussion}
Our work offer important new evidence on the role of hypotheses in the debugging process. The observations of professional developers in the preliminary study suggests that even for developers with access to shared knowledge on the Internet, finding a correct hypothesis can still be a hard and slow process. We found, in a lab setting, that developers with the correct hypothesis early in the debugging process were five times more likely to succeed. But we also found that hypotheses were few in number.  Developers were only able to formulate a median of two hypotheses, and less than half were able to formulate a correct hypothesis early in debugging. 
Even tools that suggest fault locations did not help developers to formulate more correct hypotheses or debug more successfully. 
This suggests that, while central to debugging success, formulating a correct hypothesis is often hard. 

One way to address this issue may be through creating new forms of debugging tools which more effectively surface relevant debugging hypotheses to the developer. Our study found initial indications of the potential benefits these tools might offer. Given potential explanations of the cause of a defect, developers were over six times more likely to succeed in fixing the defect. Several systems have begun to explore similar ideas in the context of helping novices diagnose programming errors. For example, HelpMeOut \cite{hartmann2010} and NoFAQ \cite{Antoni2017} suggests potential defect explanations along with fixes by identifying prior defect fixes with similar source code, error messages, or failing tests \cite{Glassman2016}. 
However, despite their promise, existing techniques for generating hypotheses are limited in ways that make them difficult to apply outside of an educational context.  These techniques rely on building a program specific database of potential defects and corresponding fixes, limiting their applicability to situations where students solve the same programming problem. 

Future debugging tools may extend the this by offering developers potential hypotheses taken from similar incorrect behavior across many different programs. But where would potential hypotheses come from? Developers gain expertise by encountering and debugging many defects. Unfortunately, developers' mechanisms for broadcasting this knowledge is limited. Developers may answers questions asked by coworkers or in the developers' community or write technical blog posts. But, as we found in our preliminary study, accessing the appropriate hypothesis for the defect at hand may remain hard. We argue that debugging tools can and should do better in helping developers share debugging knowledge. For example, debugging tools might be aware that a developer has succeeded in fixing a defect, and might prompt the developer to share a generalized explanation of what underlying cause. Other developers who face similar defects in the future might then be furnished these debugging hypotheses by the debugging tool. 

To illustrate how a developer might use such a debugging tool, consider a hypothetical example. When a developer encounters a defect, she might ask the debugging tool for a list of potential hypotheses that match the current defect context. The debugging tool could then identify relevant hypotheses, if available, and support the developer in testing them. When there are no relevant hypotheses, the developer could debug to find the  cause of the defect. When she succeeds in debugging, the debugging tool might prompts the developer to add contextual information about the defect and the correct hypothesis to a database.


\section{Conclusions}
Our studies offer new evidence for the importance of hypotheses to debugging. Lacking a correct hypothesis can be an important barrier to making progress in debugging. To address this, we found that offering a debugging aid in the form of potential hypotheses made developers six time more likely to succeed in debugging. This suggests important new opportunities for future tools that help developers to get started in finding a debugging hypothesis relevant to their task at hand.


